\documentclass{amsart}
\usepackage{graphicx}

\theoremstyle{definition}

\theoremstyle{remark}

\numberwithin{equation}{section}

\begin{document}

\title{Modeling the behavior of signal-to-noise ratio for repeated `snapshot' imaging}

\author{Junhui Li$^1$}
\address{$^1$ State Key Laboratory of Advanced Optical Communication Systems and Networks, School of Electronics Engineering and Computer Science, and Center for Quantum Information Technology, Peking University, Beijing 100871, China}
\thanks{The first author was supported in part by NSF Grant \#000000.}

\author{Bin Luo$^2$}
\address{$^2$ State Key Laboratory of Information Photonics and Optical Communications, Beijing University of Posts and Telecommunications, Beijing 100876, China}

\author{Dongyue Yang$^3$}
\address{$^3$ School of Electronic Engineering, Beijing University of Posts and Telecommunications, Beijing 100876, China}

\author{Guohua Wu$^3$}

\author{Longfei Yin$^1$}

\author{Hong Guo$^1$}
\email{hongguo@pku.edu.cn}


\keywords{Noise in imaging systems; Information theoretical analysis; Coherence imaging; Ghost imaging.}

\begin{abstract}
For imaging of static object by the means of sequential repeated independent measurements, a theoretical modeling of signal-to-noise ratio (SNR)'s behavior with varying number of measurement is developed, based on the information capacity of optical imaging systems. Experimental veritification of imaging using pseudo-thermal light source is implemented, for both the direct average of multiple measurements, and the image reconstructed by second order fluctuation correlation (SFC) which is closely related to ghost imaging. Successful curve fitting of data measured under different conditions verifies the model. 
\end{abstract}

\maketitle

In order to obtain a high quality image of a static object with low luminance, extension of the total measurement time is the usual solution, which can also depress the influence induced by short-period fluctuations on the light path. However, long exposure time imaging puts forward high requirements on the stability, noise level, and, in particular the dynamic range of the detector. Since a typical digital camera usually provides only 8 bits (256 levels) gray scale, long exposure time will easily lead to saturation for the pixels in the bright part while other pixels remain too dark to get enough illumination to form effective counts, thus lose a lot information in vain and suffer a low signal-to-noise ratio (SNR) (see, e.g., \cite{HDRbook} pp. 1). As an alternative, imaging method using repeated `snapshot' measurements with duration short enough to avoid any pixel to be overexposed during each measurement, and total number large enough to accumulate effective luminance for the less-exposed parts to be recognized, is introduced instead of a single measurement with long time exposure, and has been proven to be useful to maintain a high SNR using detectors of relatively small dynamic range (\cite{HDRbook} pp. 148). It is not difficult to imagine that the imaging SNR grows with increasing measurement number. At the same time, there will be an upper limit of SNR, considering the intrinsic noise level and imperfections of the imaging system, which means the imaging quality will not improve anymore even taking more snapshots. Naturally, one will expect a tradeoff between the measurement number and SNR. Theoretical modeling and quantitative analysis of SNR's behavior versus measurement number will provide much convenience for practical applications, e.g., prediction of ultimate SNR, and optimization of necessary measurement numbers. However, no such modeling has been presented so far. 

On the other hand, ghost imaging (GI), which enables spatial-resolved image reconstruction with correlation of a single-pixel `bucket' signal and the contactless spatial record of,  either the other split beam from the same light source \cite{Shih95}, or the spatial modulation pattern applied to the illumination beam \cite{Shapiro08}. Neither the bucket signal nor the contactless spatial record contains full information of the object's spatial distribution alone, yet the image can be reconstructed with the two parts in combine, which is the very meaning of the name `ghost'. GI is essentially a repeated snapshot imaging process, and suffers from the huge number of repeated measurements required to achieve acceptable SNR. Though the compressive sensing technique could use fewer measurements with the cost of increasing calculations \cite{Bromberg09}, the number of measurement is still too high for practical applications. While different aspects have been studied on SNR of GI \cite{Shapiro09,Lugiato10,SNR11}, quantitative analysis of the relation between SNR and the number of measurement is yet to be done. 

Despite the argument on the underlying physics \cite{Shapiro12,Shih12}, it is well accepted that the second order correlation plays an essential part in GI. Considering the identity of the two split beams used in traditional GI, second order auto-correlation of one beam should share the same merit of the cross correlation between the split beams from the same light source. Although the two characteristic properties of GI -- `ghost' and super resolution -- have been demonstrated by one-beam auto-correlation experiments \cite{SingleArm07,Kim13}, which shows that analysis on the second order auto-correlation should also apply to cross-correlation based GI, it has been pointed out that only the second order fluctuation correlation shares the exact same mechanism with GI \cite{Jane14}. 

In this Letter, we model the behavior of SNR for repeated `snapshot' measurement from the perspective of information capacity of imaging systems. Rather than the commonly used ‘invariance theorem’ \cite{Lukosz,Cox86}, phenomenological modifications inspired by Boltzmann entropy are made, corresponding to the specific process we investigate. Similar experiment setup with \cite{Kim13} is implemented to test the model, using commercial CMOS camera to capture the snapshots of static object illuminated by pseudo-thermal light. Successful fitting of measured SNR behavior validates our model, not only for direct image, but also for the image reconstructed by the second order fluctuation correlation, which suggests the model's capability for GI. 

The information capacity for imaging system originates from the time-bandwidth product (sampling time $T$ times bandwidth $B_T$) in signal processing theory, and its correspondence in spatial domain –-  space-bandwidth product (sampling length $L_i$ times spatial bandwidth in that direction $B_i$, $i = x, y$) in optical imaging theory (see, e.g., \cite{GoodmanBook} pp. 27). They represent the minimum number of samples required for the signal to be properly sampled in temporal/spatial domain. These two kinds of products were originally believed to be independently invariant for measurements under different parameter settings using the same 2D imaging system, however, later study on superresolution imaging shows that it is, in essence, the number of degrees of freedom $N_F = 2\left( 1 + 2L_xB_x \right) \left( 1 + 2L_yB_y \right) \left( 1 + 2TB_T \right)$, the combination of temporal and spatial products, that is a constant \cite{Lukosz, Cox86}. Later on, the detected noise in imaging system is taken into account, and a new invariant named `information capacity' $C$ is introduced, namely,\cite{Cox86}
\begin{equation}
\begin{aligned} 
C &= \log {\left[ {{{\left( {\frac{{s + n}}{n}} \right)}^{{1 \mathord{\left/
 {\vphantom {1 2}} \right.
 \kern-\nulldelimiterspace} 2}}}} \right]^{{N_F}}}\\
 &= \left( {1 + 2T{B_T}} \right)\prod\limits_{i = x,y} {\left( {1 + 2{L_i}{B_i}} \right)}  \times \log \left( {1 + \rm{SNR}} \right),
\end{aligned}
\label{eq:refname1}
\end{equation}
where $s$ and $n$ represent the intensity of the detected signal and noise, respectively, and the signal-to-noise ratio SNR $ = {s \mathord{\left/ {\vphantom {s n}} \right. \kern-\nulldelimiterspace} n}$. $C$ should be invariant for any one-time measurement under different parameter settings conducted by a particular imaging system. 

Instead of applying in repeated snapshot measurement process directly, the underlying significance of Eq. (\ref{eq:refname1}) should be examined first. In fact, $C$ can be seen as an analogue of Boltzmann entropy $S = k_B \log{ \Omega}$ in thermal dynamics, which is proportional to the logarithm of $\Omega$, the total number of possible states. Square root of the intensity ratio $ {{\left( {s + n} \right)} \mathord{\left/ {\vphantom {{\left( {s + n} \right)} n}} \right. \kern-\nulldelimiterspace} n}$ is the average number of states the measurement can distinguish at one sample point, in the sense of the joint space-time, with $N_F$ being the total number of sample points. For an imaging process, the information capacity $C$ tells the amount of the object’s unknown information after the image has been generated. In other words, $C$ indicates, besides the information revealed by the image, the amount of the uncertainty the system can hold after the imaging process. 

As for the repeated snapshot measurement of a static object under stationary and ergodic light source, the situation is slightly different. The final reconstructed image is built on sequential repeated measurements with identical parameter setting. Each measurement reveals some information about the object, and the system’s uncertainty is reduced. That is to say, instead of being invariant, the information capacity decreases with the growing number of involving measurements (see, e.g., \cite{Bromberg09}). If the time separation between the neighboring measurements is larger than the coherence time of the light source, which is often the case, and that since the same static object is illuminated by a stationary and ergodic light, the information of the object revealed by different measurements can be seen as independent identically distributed (i.i.d.). Then, in the sense of average, contribution from each measurement will cost the same amount of decrease in information capacity $C$. Therefore, for a repeated snapshot measurement built by $N$ measurements, each containing information $m$ in average, the information capacity $C$ of the system becomes
\begin{equation}
C = C_0 - Nm,
\label{eq:refname2}
\end{equation}
where $C_0$ is the informational uncertainty of the system before the first measurement. 

The term $\left(1 + \rm{SNR} \right) $ in Eq. (\ref{eq:refname1}) should be reconsidered for the repeated snapshot measurement process. This part should be the number of possible different states which can be distinguished by the system at one sampling point, or to say, the unknown information of the object which has not been revealed by the image yet. As for an imaging process, it has been known that the rightful form of SNR is different from the usual way used in signal processing \cite{SNR02}. Therefore, the meaning of `signal' and `noise' should be redefined. A usual way is to take the image plane average of an ideal image of the object $O\left(x,y\right)$ ($x$, $y$ are coordinates of the image plane) as the `signal', which stands for a measure of the amount of information contained in the imaging process, and treat the standard derivation of the actual reconstructed image $R \left( x,y \right)$ from $O \left( x,y \right)$ as an estimation of the `noise'.  In this way, SNR reads
\begin{equation}
{\rm{SNR}} = \frac{{{{\left\langle {O\left( {x,y} \right)} \right\rangle }_{\left( {x,y} \right)}}}}{{\left\langle {{{\left( {R\left( {x,y} \right) - O\left( {x,y} \right)} \right)}^2}} \right\rangle _{\left( {x,y} \right)}^{{1 \mathord{\left/
 {\vphantom {1 2}} \right.
 \kern-\nulldelimiterspace} 2}}}},
\label{eq:refname3}
\end{equation}
which stands for the number of states distinguishable due to the variance of the reconstructed image differing from the object. Unlike that in Eq. (\ref{eq:refname1}), SNR in Eq. (\ref{eq:refname3}) gives the amount of the system’s known information revealed by the image. Therefore, the number of possible states at one sampling point unrevealed in the reconstructed image should have the form of $\left( \Omega_0 - \rm{SNR} \right)$, in which $\Omega_0$ stands for the number of possible states before imaging. If the measurement parameters of the spatial domain are unchanged for all the measurements, the spatial part of Eq. (\ref{eq:refname1}) can be seen as a constant, i.e., $\left( 1 + 2L_xB_x \right) \left( 1 + 2L_yB_y \right) = {1 \mathord{\left/ {\vphantom {1 A}} \right. \kern-\nulldelimiterspace} A}$. Substituting Eq. (\ref{eq:refname2}) into Eq. (\ref{eq:refname1}) with $\left( \Omega_0 - \rm{SNR} \right)$ replacing $\left( 1 + \rm{SNR} \right)$ yields
\begin{equation}
{\rm{SNR}} = \Omega _0 - {e^{{C_0}}}\exp \left( { - \frac{{Am}}{{1 + 2T{B_T}}}N} \right).
\label{eq:refname4}
\end{equation}
Eq. (\ref{eq:refname4}) indicates that SNR evolves negative-exponentially with the growth of the measurement number $N$. It converges to an upper limit $\Omega_0$ when $N \to \infty$, and all the system’s information have been revealed. The growing speed of SNR is ever-decreasing, with a converge constant $k = {{\left( {2T{B_T} + 1} \right)} \mathord{\left/ {\vphantom {{\left( {2T{B_T} + 1} \right)} {\left( {Am} \right)}}} \right. \kern-\nulldelimiterspace} {\left( {Am} \right)}}$, the number of measurement required for SNR to fulfill ${1 \mathord{\left/ {\vphantom {1 e}} \right. \kern-\nulldelimiterspace} e}$ of the total growth from $0$ to $\Omega_0$, denoting the speed of converge. 

\begin{figure}[htbp]
\centering
\includegraphics[width=\linewidth]{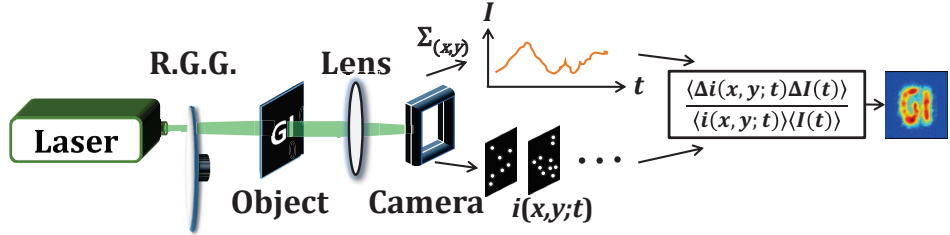}
\caption{Experiment setup. Pseudo-thermal light generated by laser passing rotating ground glass (R.G.G.) illuminates static object `GI', then projected onto CMOS camera by lens. Direct image is the average of all snapshots. The second order fluctuation auto-correlation is calculated between each snapshot and its corresponding `bucket' -- the total count of that snapshot.}
\label{fig:1}
\end{figure}

An experiment is implemented to test the model. As shown in Figure \ref{fig:1}, light from a $532~\rm{nm}$ semiconductor laser (Ceo DPSSL-532U, with power less than $32~\rm{mW}$ and linewidth less than $0.1~\rm{MHz}$) passing through a $4~\rm{rpm}$ rotating ground glass plate (Edmund 100 mm diameter $220$ grit) makes a pseudo thermal source with about $30~\rm{ms}$ coherence time (which is smaller than $40~\rm{ms}$ -- the time separation between neighboring measurements, to make sure that the snapshots are independent). This pseudo-thermal light illuminates a static `GI' pattern object of $3~\rm{mm}$ square size (which is much bigger than the coherence length of the light source on the object plane -- $0.02~\rm{mm}$). A imaging lens with $100~\rm{mm}$ focus length forms an image on a commercial 8-bit CMOS camera (Thorlabs DCC 3240C, minimum exposure $0.009~\rm{ms}$), illuminates about $100 \times 100$ pixels. Within each short duration snapshot, when the laser outputs its maximum power, speckles generated by the ground glass grits can be recognized clearly from the image, which suggests the existence of short-period fluctuations. Repeated snapshots under different exposure time and laser power are recorded. Direct image is simply the average of all the snapshots. Image reconstructed by the second order fluctuation auto-correlation is calculated from
\begin{equation}
R\left( {x,y} \right) \propto \frac{{{{\left\langle {\left[ {i\left( {x,y;t} \right) - {{\left\langle {i\left( {x,y;t} \right)} \right\rangle }_t}} \right] \times \left[ {I\left( t \right) - {{\left\langle {I\left( t \right)} \right\rangle }_t}} \right]} \right\rangle }_t}}}{{{{\left\langle {i\left( {x,y;t} \right)} \right\rangle }_t}{{\left\langle {I\left( t \right)} \right\rangle }_t}}},
\label{eq:refname5}
\end{equation}
where $i \left( x,y;t \right)$ and $I \left( t \right) = \sum\nolimits_{x,y} {i\left( {x,y;t} \right)} $ are the spatial record and total count of the snapshot captured at time $t$, respectively. The area of interest (AOI) of each snapshot is set to be $160 \times 160$ pixels with fixed coordinates, larger than the size of the image in the center. The ground glass is taken away to get an ideal image without fluctuation from direct illumination of high power laser under sufficiently long exposure time. After setting a threshold to binarize the ideal image, then being adjusted into the same size of $R \left( x,y \right)$, and being aligned with the center of $R \left( x,y \right)$, SNR is calculated by Eq. (\ref{eq:refname3}).  

\begin{figure}[htbp]
\centering
\includegraphics[width=\linewidth]{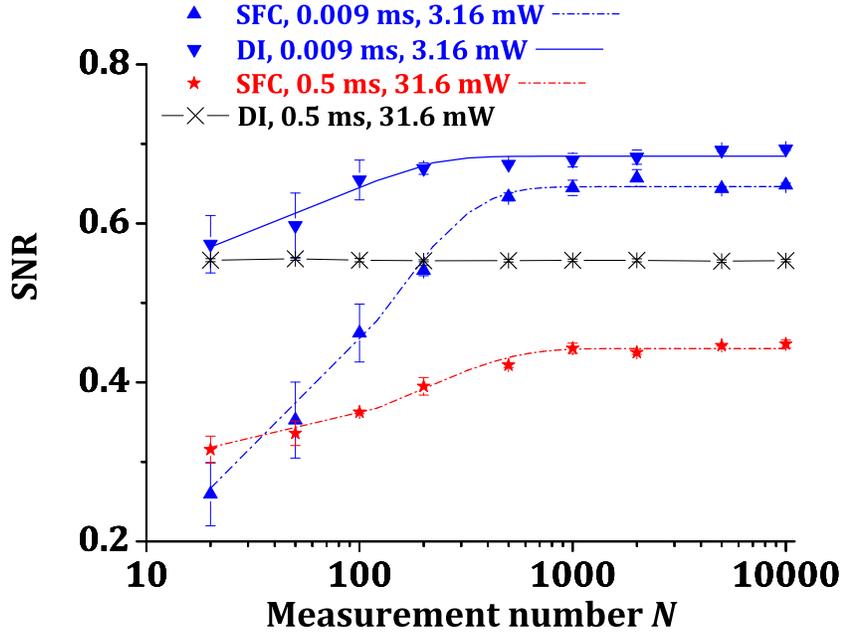}
\caption{Measured SNR versus measurement number $N$. `SFC' means the second order fluctuation correlation, and `DI' is short for direct imaging. The points stand for the measured data, and the lines are fitting curves by Eq. (\ref{eq:refname4}), except the black solid line with cross marks. }
\label{fig:2}
\end{figure}

\begin{table}[htbp]
\centering
\caption{\bf Fitting Results for SNR vs. measurement number $N$ by Eq. (\ref{eq:refname3})}
\begin{tabular}{cccc}
\hline
Exposure/ms & Power/mW & Type & Adjusted $R^2$ \\
\hline
$0.009$ & $3.16$ & SFC & $0.99449$ \\
$0.009$ & $3.16$ & DI & $0.95435$ \\
$0.5$ & $31.6$ & SFC & $0.98775$ \\
\hline
\end{tabular}
  \label{tab:1}
\end{table}

Measured SNR upon varying measurement number $N$ are shown in Figure \ref{fig:2}, both for direct imaging (DI) and second order fluctuation auto-correlation (SFC). While data under other settings are also collected but not shown, only the two extreme condition with respect to two parameters, i.e., light power and exposure time, are given in Figure \ref{fig:2}. They all fit Eq. (\ref{eq:refname4}) well, as shown in Table \ref{tab:1}, that each has a close-to-unity adjusted $R$ squared coefficient (see, e.g., \cite{R2}). The range of exposure time is more than $50$, and the large light power is about $10$ times of the small one. Therefore, as for the average number of photons registered by the camera during each snapshot, our model validates for the range of at least two and a half magnitudes, for both direct imaging and image reconstructed by the second order fluctuation correlation. 

The only exception is the SNR of direct imaging in the strong luminance case ($31.6~\rm{mW}$ bright laser illumination under $0.5~\rm{ms}$ long exposure), which seems to be a constant with growing $N$. It suggests that the converging process finishes within $20$ snapshots, which is the smallest measurement number in our experiment. On the other hand, relatively large error bars when $N$ is small (less than $100$) for the weak luminance case shows large fluctuations exist because there has not been sufficiently many photons registered to form a stable image.
 
\begin{figure}[htbp]
\centering
\includegraphics[width=\linewidth]{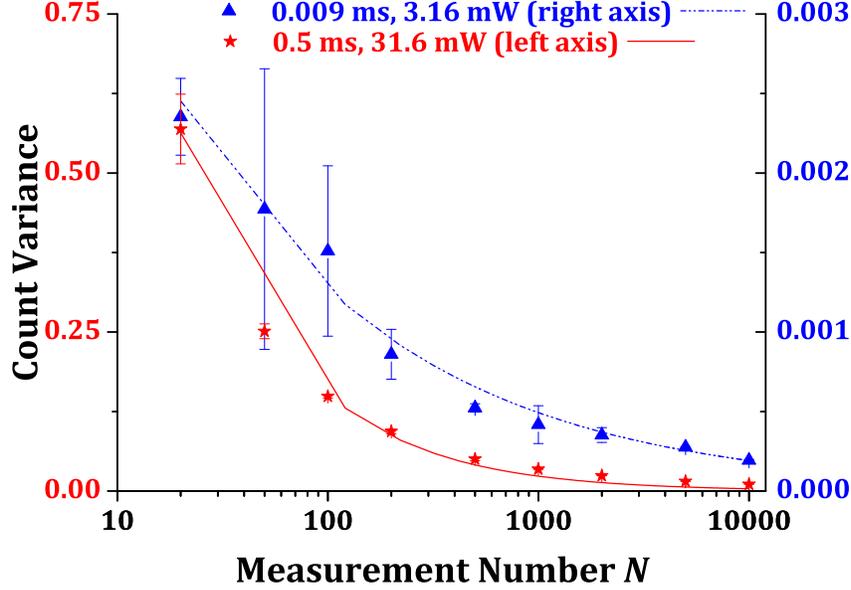}
\caption{Measured count variance $\Delta i$ versus measurement number $N$. The points stand for the measured data, and the lines are fitting curves by power function $\Delta i = a N^b$. }
\label{fig:3}
\end{figure}

The only assumption during the whole derivation of Eq. (\ref{eq:refname4}) is that the snapshots are i.i.d.. To check whether this assumption is fulfilled, total count's variance $\Delta i$ is plotted versus the measurement number $N$, as Figure \ref{fig:3} shows. 

\begin{table}[htbp]
\centering
\caption{\bf Fitting Results for count variance $\Delta i$ vs. measurement number $N$ by power function $\Delta i = a N^b$. }
\begin{tabular}{ccccc}
\hline
Exposure/ms & $a$ & $b$ & Adjusted $R^2$ \\
\hline
$0.009$ & $0.0838 \pm 0.0105$ & $-0.410 \pm 0.031$ & $0.974$ \\
$0.5$ & $6.44 \pm 0.60$ & $-0.813 \pm 0.028$ & $0.996$ \\
\hline
\end{tabular}
  \label{tab:2}
\end{table}

A power function $\Delta i = a N^b$ is used to fit the data points in Figure \ref{fig:3}. The results are demonstrated in Table \ref{tab:2}. For a classical independent identically distributed random variable, the power law $b$ should approach $-0.5$ when $N \to \infty$, as what the central limit theorem points out, while for quantum process with strong correlation, $b$ can even approach $-1$. Although the weak luminace case ($3.16~\rm{mW}$ laser under $0.009~\rm{ms}$ exposure) stays in the classical regime, the other one with high power laser and long time exposure exceeds the classical limit $b=-0.5$. This observation reveals an interesting paradox, that the $\Delta i - N$ relation seems to contradict the i.i.d assumption. Meanwhile, it is this very assumption that leads to Eq. (\ref{eq:refname4}), which shows high consistency with measured SNR. Further investigation is needed to clarify this paradox. 

It should be noted that, though fitting the measured data quite well, Eq. (\ref{eq:refname4}) is, to some extent, a phenomenological formula. The validity regime of Eq. (\ref{eq:refname4}) could be given by testifying more cases using different setups with all kinds of light sources and different setups.Investigation on each parameter's influence in Eq. (\ref{eq:refname4}) will be meaningful. Although the second order fluctuation auto-correlation in this Letter shares the same formula with real GI (Eq. (\ref{eq:refname5})), and it is fair to say our model should also be suitable for GI, experiment verification is still necessary. 

In conclusion, based on the theory of information capacity for imaging system, we study the repeated `snapshot' measurement imaging process. A model for the relation between SNR and the number of measurement is developed, which shows the increasing converge behavior of SNR with growing measurement number correctly and fits experiments of both direct imaging and second order fluctuation auto-correlation with high accuracy. This model may suit for any repeated, identical, yet independent sampling on static object. 

\subsection*{Funding}
National Science Fund for Distinguished Young Scholars of China (61225003); Natural Science Foundation of China (61471051, 61401036); Postdoctoral Science Foundation of China (2015M580008); the 863 Program; PhD Students' Overseas Research Program of Peking University, China.

\bibliographystyle{amsplain}

\end{document}